\documentclass[a4paper,american]{scrartcl}
\usepackage{lmodern}
\usepackage[utf8]{inputenc}
\usepackage[T1]{fontenc}
\usepackage{color}
\usepackage{todonotes}
\usepackage{babel}
\usepackage{amsmath}
\usepackage{amssymb}
\usepackage{setspace}
\usepackage[authoryear]{natbib}
\usepackage[font=small]{quoting}
\usepackage{latexsym}
\usepackage[bookmarks=false,pdfborder={0 0 0},colorlinks=false]{hyperref}
\usepackage[normalem]{ulem}

\makeatletter

\renewcommand{\Im}{\operatorname{Im}}

\makeatother

\begin{document}

\title{What is matter? The fundamental ontology of atomism and structural realism}

\author{Michael Esfeld%
\thanks{Université de Lausanne, Faculté des lettres, Section de philosophie,
1015 Lausanne, Switzerland. E-mail:
\protect\href{mailto:Michael-Andreas.esfeld@unil.ch}{Michael-Andreas.Esfeld@unil.ch}}%
, Dirk-André Deckert%
\thanks{Ludwig-Maximilians-Universität München, Mathematisches Institut,
Theresienstrasse 39, 80333 München, Germany. E-mail: \protect\href{mailto:deckert@math.lmu.de}{deckert@math.lmu.de}%
}, Andrea Oldofredi%
\thanks{Université de Lausanne, Faculté des lettres, Section de philosophie,
1015 Lausanne, Switzerland. E-mail: \protect\href{mailto:Andrea.Oldofredi@unil.ch}{Andrea.Oldofredi@unil.ch}}
}

\maketitle
\begin{abstract} 
    \begin{center} Draft for Anna Ijjas and Barry Loewer
        (eds.):\\ A guide to the philosophy of cosmology, Oxford University
        Press 
    \end{center}
    \medskip

    We set out a fundamental ontology of atomism in terms of
    matter points. While being most parsimonious, this ontology is able to match both classical and
    quantum mechanics, and it remains a viable option for any future
    theory of cosmology that goes beyond current quantum physics. The matter points are structurally individuated: all there is to them are the spatial relations in which they stand; neither
    a commitment to intrinsic properties nor to an absolute space is required.
    The spatial relations change. All that is needed to capture
    change is a dynamical structure, namely dynamical
    relations as expressed in terms of the dynamical parameters of a physical
    theory. \medskip

    \noindent \emph{Keywords}: atomism, ontic structural realism, primitive
    ontology, dynamical structure, relationalism, classical
    mechanics, Bohmian mechanics
    
\medskip
   
\end{abstract}

\tableofcontents{}

\section{Introduction: The need for an ontology for cosmology}
\label{sec:introduction}

Cosmology is concerned with the universe as a whole. It is therefore of central interest to cosmology to obtain answers to the following questions: 
\begin{enumerate}
    \item What is matter? What is space and time?

    \item What are the laws of nature?

    \item How does matter in space and time, being subject to certain laws, explain the observable phenomena?
\end{enumerate}

These are also the questions on which the traditional stance of natural philosophy focuses, as exemplified, for instance, in the works of Descartes, Leibniz, and Newton. In their work, physics and philosophy come together in an inseparable manner in the search for an answer to these questions.

This paper is an exercise in natural philosophy in that spirit. Our aim is to
sketch out a fundamental ontology of the natural world that can serve as a basis
for cosmology in answering these questions while being most simple and most
general. In order to achieve that aim, we take up atomism as regards matter and
relationalism as regards space and time, bringing these traditional positions
together with the stance that is known as ontic structural realism in today's
philosophy of science.

More precisely, in cosmology, it is evident that we need a quantum theory
without observers -- that is, a quantum theory that does not rely on
notions such as observables, observers, measurements and measurement apparatuses as basic,
undefined concepts, but derives an account of these from its axioms. 
This issue is encapsulated nicely in this quote from \cite{Bell:2004aa}:
\begin{quote}
    Was the wavefunction
    of the world waiting to jump for thousands of millions
    of years until a single-celled living creature appeared? Or
    did it have to wait a little longer, for some better qualified
    system . . . with a PhD? (\cite{Bell:2004aa}, p. 217) 
\end{quote}
Because of these problematic notions,
the quantum theory going back to \cite{Everett:1957aa} is widespread in
cosmology (see the contribution of Wallace to this volume): this theory admits
only the quantum state of the universe in its ontology (as represented by the
universal wave function); this state always undergoes a linear evolution (as
represented e.g. by the Schrödinger equation) (see \cite{Wallace:2012aa} for the
most detailed contemporary account). However, although it is a consistent
quantum theory without observers, for these reasons, it is also doubtful whether
this theory can achieve an account of experience (see notably
\cite{Maudlin:2010aa}) and whether it can provide a place for probabilities (see
e.g. \cite{Dawid:2014aa} and \cite{Dizadji:2015aa}).

Apart from Everettian quantum mechanics, there is another way to achieve a
quantum theory without observers available that takes a completely different
direction: instead of starting from the quantum state and regarding the quantum
state as the referent of the formalism, this approach bases itself on the claim
that a fundamental ontology of matter arranged in three-dimensional space or
four-dimensional space-time is called for in quantum physics like in classical
physics in order to achieve an account of experience; probabilities then are
probabilities of the behaviour of the elements of that ontology. The claim thus
is that what is known as a \emph{primitive ontology} of a configuration of
matter in ordinary space is a necessary condition to account for the observable
phenomena (the term ``primitive ontology'' goes back to \cite{Durr:2013aa}, ch.
2, originally published 1992; cf. also Bell's notion of local beables in
\cite{Bell:2004aa}, ch. 7); to turn that necessary condition into a sufficient
one, one has to formulate a dynamics for that configuration that excludes
superpositions of matter in space, but has to include entanglement. The quantum
state and its evolution then comes in as the appropriate means to perform that
task: the quantum state is not a physical object, but a dynamical variable whose
job it is to represent the evolution of the distribution of matter in ordinary
space. We adopt this stance in this paper, relying on its elaboration in the
recent literature (see notably \cite{Allori:2008aa}, \cite{Esfeld:2014ab,
Esfeld:2015aa}, \cite{Maudlin:2015aa}; see also the contribution by Goldstein,
Struyve and Tumulka to this volume). Our aim is to show that such an ontology is
a suitable framework for cosmology, doing in fact a better job than the
Everettian framework of admitting only the quantum state.

If a primitive ontology of a configuration of matter in ordinary space is called for in quantum physics as in classical physics, the best starting point to introduce this ontology is atomism, which is also the oldest and most influential tradition in natural philosophy,
going back to the pre-Socratic philosophers Leucippus and Democritus. The latter
is reported as maintaining that \begin{quote} ... substances infinite in number
    and indestructible, and moreover without action or affection, travel
    scattered about in the void. When they encounter each other, collide, or
    become entangled, collections of them appear as water or fire, plant or man.
    (fragment Diels-Kranz 68 A57, quoted from \cite{Graham:2010aa}, p. 537)\end{quote}

In a similar vein, Newton writes at the end of the Opticks: \begin{quote} ... it
    seems probable to me, that God in the Beginning form'd Matter in solid,
    massy, hard, impenetrable, moveable Particles ... the Changes of corporeal
    Things are to be placed only in the various Separations and new Associations
    and motions of these permanent Particles. (\cite{Newton:1952aa}, question 31, p.
    400)\end{quote}

The attractiveness of atomism for our endeavour is evident from these quotations: on the one hand, it is a proposal
for a fundamental ontology that is parsimonious and that is general in that it covers everything, from microscopic objects to entire galaxies. On the other hand, it offers a clear and simple explanation of the realm of our
experience. Macroscopic objects up to entire galaxies are composed of fundamental, indivisible
particles. All the differences between the macroscopic objects
-- at a time as well as in time -- are accounted for in
terms of the spatial configuration of these particles and its change, which is subject to certain laws.

Whereas atomism is a purely philosophical proposal in Leucippus and Democritus, it is turned into a precise physical theory by Newton. Accordingly, 
classical mechanics -- and classical statistical physics -- are usually seen as the
greatest triumph of atomism. However, the impression is widespread that atomism is outdated by the physics of the 20th century. We, by contrast, think that such a judgement is premature, mainly for the following three reasons:
\begin{enumerate}
    \item All experimental evidence in fundamental physics is evidence of
        particles -- from dots on a display to traces in a cloud chamber.
        Entities that are not particles -- such as waves or fields -- come in as
        figuring in the explanation of the behaviour of the particles, but they
    are not themselves part of the experimental evidence: an electric field
        is probed by the behaviour of a test charge subject to it, the outcome of the double
        slit experiment consists in particles hitting on a screen, etc.

    \item The most elaborate proposal for a quantum theory without observers that relies on a primitive ontology of matter distributed in ordinary space is the one going back to \cite{Broglie:1928aa} and \cite{Bohm:1952aa} and advocated by \cite{Bell:2004aa}. Its dominant contemporary version is known as Bohmian mechanics (see \cite{Durr:2013aa} and the textbook \cite{Durr:2009fk}). The fundamental or primitive ontology that this theory proposes is the same as the one of classical mechanics, namely a configuration of point particles in ordinary space moving on continuous trajectories. What changes is the law that describes this motion and the dynamical parameters that figure in this law. We'll consider Bohmian mechanics at the end of section~\ref{sec:dynamical} and will also briefly mention other primitive ontology theories of quantum mechanics there.   

    \item Going beyond quantum mechanics, there is to this day no fully worked
        out ontology for quantum field theory available, not to mention quantum
        gravity. In part, this is due to the fact that quantum field theory is
        plagued by the notorious occurrence of mathematical infinite and thus
        ill-defined quantities.  \cite{Dirac:1963aa} famously referred to the
        missing ontology and the lack of a rigorous mathematical understanding
        as \emph{class one} and \emph{class two} difficulties, hoping that
        progress with respect to the latter will clear up the former. This hope
        has not yet materialized: we neither dispose of an elaborate ontology of
        this domain that replaces the commitment to particles with something
        else in explaining the experimental evidence, nor of an elaborate
        particle ontology. When introducing mathematical cut-offs to render
        otherwise infinite quantities finite, versions of a particle ontology
        have been attempted notably by \cite{Durr:2005aa} and
        \cite{Durr:2013aa}, ch. 10, as well as \cite{Colin:2007aa} and
        \cite{Deckert:2010aa}, chs. 6 and 7, based on \cite{Dirac:1934}. In any
        case, the measurement problem plagues quantum field theory in the same
        way as it concerns quantum mechanics (see e.g. \cite{Barrett:2014aa}),
        and the argument for a primitive ontology applies to any quantum theory,
        including quantum field theory (see e.g. \cite{Bell:2004aa}, ch. 19),
        and quantum gravity (see notably \cite{Goldstein:2001aa} and
        \cite{Vassallo:2014aa}). That is to say, there is reason to assume that
        the scheme that we set out in this paper will also work for quantum
        field theory: a primitive ontology of permanent point particles that are
        characterized by the spatial relations in which they stand and a
        dynamical structure that captures the change in these spatial relations,
        with fields belonging to that dynamical structure instead of being
        primitive objects on a par with (or instead of) the particles. 
        Such an ontology must however be compatible with the
        notions of
        creation, annihilation, and number operators of quantum field theory.
        We'll
        briefly explain this view of fields in section~\ref{sec:spatial and dynamical}. 

\end{enumerate}

In sum, for these reasons, we think that trying out the old and venerable paradigm of atomism still is a worth while enterprise. In this paper, we show how this paradigm can profit from the contemporary stance of ontic structural realism in order to be vindicated as a proposal for a fundamental ontology that is most parsimonious, while being able to explain the observable phenomena, including serving as a basis for cosmology. Thus, in the next section, we set out a fundamental ontology of matter points in terms of spatial structure: all there is to the matter points is their spatial configuration, that is, the network of the spatial relations in
which they stand. Since these relations change, a dynamical structure applies to
the particle configuration. In section~\ref{sec:dynamical}, we
introduce the dynamical structure; in section~\ref{sec:spatial and dynamical}, we discuss its nature.

\section{Spatial structure: The fundamental ontology of matter points}
\label{sec:spatial}

Both Democritus and Newton conceive the atoms as being equipped with a few basic
intrinsic properties -- that is, properties that each matter point has taken
individually, independently of whether or not there are other atoms. The
paradigmatic example is mass in Newtonian mechanics. However, also in Newtonian
mechanics, both inertial and gravitational mass are introduced through their dynamical role, namely as dynamical parameters that couple the motions of the particles to one another, as was
pointed out by \cite{Mach:1919aa}, p. 241, among others: in short, mass tells us
something about how the particles move. Hence, the role of mass is the one of a
parameter that expresses a dynamical relation among the atoms. The same argument
as for gravitational mass applies to charge. Nonetheless, it is possible to maintain that parameters such as mass and charge are introduced in classical physics through the dynamical role that they
play for the motion of the particles, but that the description in terms of a
dynamical role refers to an underlying intrinsic property, since these
parameters are attributed to the particles taken individually (see e.g.
\cite{Jackson:1998aa}, pp. 23-24).

In any case, when it comes to quantum mechanics, further doubt is cast on the
conception of intrinsic properties: at first glance, also quantum mechanics
looks like attributing values of mass and other properties to the particles
taken individually. However, experimental considerations involving interference
phenomena -- for instance in the context of the Aharonov-Bohm effect and of
certain interferometry experiments -- show, in brief, that mass and charge are
effective at all the possible particle locations that the quantum state admits
(see \cite{Brown:1995aa, Brown:1996aa} and references therein as well as
\cite{Esfeld:2015aa} for a detailed assessment). Consequently, these parameters
are situated on the level of the quantum state of the particles, which is
represented by their wave function; given entanglement, it is not possible to
attribute a quantum state to the particles taken individually, but only to their
configuration as a whole. Therefore, quantum entanglement -- but also already
the mentioned classical considerations concerning mass and charge -- are a central
motivation to abandon the view of physical objects being characterized by
intrinsic properties and to replace this view with \emph{structural realism}, which
lays stress on relations. We will go into structural realism below in this section.

If mass and charge, as well as the quantum state, express in fact dynamical
relations among the particles that tell us something about their motion, then
what remains as basic characteristic of the atoms is their position in space,
given that motion is change in the spatial arrangement of the particles. In
fact, Democritus and Newton conceive the atoms as being inserted into an
absolute background space, which is three-dimensional and Euclidean. However,
this commitment is problematic as well. If there is an absolute space, then
there are real differences that do not make a physical difference: there are
many different possibilities to place the whole configuration of matter in
absolute space that leave the spatial relations among the atoms unchanged so
that there is no physical difference between them, as Leibniz points out in his
famous objections to Newton's view of space and time (see notably Leibniz' third
letter, §§ 5-6, and fourth letter, § 15, in \cite{Leibniz:1890aa}, pp. 363-364,
373-374, English translation \cite{Leibniz:2000aa}). Moreover, this view faces
the following question: What is it that occupies points of space? In other
words: What accounts for the difference between a point of absolute space being
occupied and its being empty? If there is a dualism of space and matter, one has
to answer the question of what distinguishes matter from space; but relying on
intrinsic properties that establish such a distinction is problematic (as are the
conceptions of the atoms being bare -- that is, propertyless -- substrata, or
them having a primitive thisness -- that is a qualitative feature of each atom
of being that atom that is independent of all its properties; these metaphysical
conceptions have no basis in physics).

These considerations are a further good motivation for going structural, that
is, to follow Leibniz in abandoning the dualism of matter and space and
switching to relationalism about space. Thus, in the following, by
\emph{position}, we mean the spatial \emph{relations} among the matter points,
making up their configuration. In an Euclidean conception of space, the spatial
relations can be described in terms of the corresponding Euclidean lengths and
directions in terms of angles. For more complicated geometries (such as
Riemannian geometry), this notion of directions and lengths has to be adapted
appropriately, which bares no principle obstacle. If change can be
described as a parametrized list of configurations of spatial relations, then velocity is
the first derivative of these \emph{relations} with respect to an appropriate
parametrization, i.e., time. Consequently, the fundamental ontology is
\emph{background independent}: there are only the spatial relations among the
matter points, but no physical background space. We
shall refer to the network of spatial relations connecting the matter points as
the \emph{spatial structure}.

This structure consists only in what is known as \emph{Leibnizian relations}
(see \cite{Maudlin:1993aa}, p. 187), namely the spatial relations among matter
points at a time and their temporal derivative. Consequently, relationalism
counts all transformations that leave the spatial relations between the matter
points within a given configuration unchanged as mathematical representations of
one and the same physical configuration. In other words, there is a difference
between configurations of matter points only if there is motion of some atoms
relative to other atoms in the universe. Relationalism thereby avoids the
commitment to differences in the ontology that do not amount to a physical
difference.

If there is no absolute space into which the matter points are embedded, there
is no geometry fixed from the outset in the following sense: the spatial
relations in any given initial configuration of matter points do not determine
the geometry of the space in which the temporal development of these relations
can be represented as being embedded. For instance, it may be the case that the
spatial relations in a given initial configuration of matter points can be
described in terms of Euclidean lengths and angles, but that the relations then
develop in such a way that they can no longer be described in Euclidean terms (so that the overall
geometrical space in which these relations can be represented as being embedded
is not Euclidean). In short, the Leibnizian relations \emph{at a time}
are insufficient to fix the geometry. But the Leibnizian relations \emph{as a
whole} are sufficient to do so: given the entire change in the spatial relations
in the configuration of matter points of the universe, the geometry best suited to
describe the universe is fixed by the history of these relations as a whole (see
\cite{Huggett:2006aa}, pp. 50-55). That is to say, geometry and dynamics come in a package, instead of the geometry being given from the outset and constraining the
dynamics. This view is thereby wide
enough to leave the possibility open that a future theory of quantum cosmology
and gravity may treat space as emergent: all that is required is that the basic relations \emph{evolve} in such a way that they
manifest a certain geometry; this is what makes them spatial relations.

Since there is no geometry given from the outset, relationalism has to
presuppose all the spatial relations that connect the points in a given initial
configuration of matter points of the universe as a primitive (cf. the
objections that \cite{Field:1985aa}, pp. 49-53, and \cite{Maudlin:2007aa}, pp.
87-89, build on this fact). However, as far as the ontology is concerned, this
is in fact less than what the substantivalist about space has to presuppose: the adherent to the view of space being a physical object has to presuppose all the points of space and all the facts of which of these
points are occupied by material entities as primitive. The geometry of space
enables the substantivalist to derive the spatial relations among matter points
from the geometry of space, but this does not make their ontology simpler. To
the contrary, subscribing to the existence of an absolute space that stretches
out to infinity far beyond where any matter exists in order to describe a large,
but finite configuration of point particles that are spatially related can with
good reason be dismissed as a rather extravagant commitment. In sum, the main
argument for relationalism about space is that it is the more parsimonious
ontology than substantivalism: it avoids a commitment to a dualism of matter and
space, and it eschews a commitment to real differences in the spatial
arrangement of matter that do not make a physical difference.

Endorsing relationalism about space enables us to answer the question of what is
matter without having to bring in intrinsic features: \emph{there are particles,
namely matter points. These are matter points, because there is a
non-vanishing spatial relation between any two such points.} In a nutshell,
the essence of matter is that material objects stand in spatial relations.
If one is committed to a dualism of matter and space and characterizes
matter points in terms of their position in space, it is the points of space
that are individuated by the topology and the geometry of space.
Consequently, one runs into the impasse of having to answer the question of
what constitutes the difference between an unoccupied and an occupied point
of space. By contrast, if there are only matter points, but no points of
space, then the spatial relations among these points are available to answer
the question of what characterizes matter. This characterization furthermore
accounts for the impenetrability of matter without having to invoke a notion
of mass: matter points are impenetrable because for there to be two matter points, there has to be a non-vanishing
relation of spatial distance between them. That is why spatial relations are suitable to characterize what exists together, as Leibniz stresses by his definition of space as the order of coexistences (see notably Leibniz' third letter, §
4, in \cite{Leibniz:1890aa}, p. 363, English
translation \cite{Leibniz:2000aa}): the spatial relations make it that there is a numerical plurality of objects (matter points), because they are able to establish a distinction among the matter points (provided that their configuration is not symmetric). 

We thus take up the Cartesian conception of matter in
terms of spatial extension -- \emph{res extensa} or extended substance --, replacing the scholastic one in terms of (occult) qualities, but propose to conceive extension in terms of spatial relations that connect unextended points: what accounts for something being a matter point -- by contrast, say, to a hypothetical mind point -- is its standing in spatial relations to other such points. Thus, there are no points of space or space-time in the world. There are objects that are not extended in themselves (points). These are material -- namely matter points --, because they stand in spatial relations. If they were not standing in spatial relations -- but, say, in hypothetical fundamental thinking relations --, they would not be matter and not be physical entities, but mental entities.

Consequently, the matter points are not bare substrata. They are structurally individuated, namely by
spatial relations: all there is to the matter points are the spatial relations
in which they stand, making up a spatial configuration. As the literature on
ontic structural realism has made clear, structures in the sense of concrete
physical relations -- such as spatial relations -- can individuate physical objects (see e.g. \cite{Ladyman:2007aa}). We advocate here a \emph{moderate} ontic structural realism (see \cite{Esfeld:2004aa, Esfeld:2011aa}): there are fundamental physical objects, namely matter points; but all there is to these objects are the spatial relations among them. Thus, they do not have an intrinsic nature, but a relational one.

We do not see a physical or a metaphysical reason to
conceive ontic structural realism as being opposed to an object-oriented
metaphysics (see by contrast \cite{Ladyman:2007b} and \cite{French:2014aa}): if there are relations,
there are objects that stand in the relations. In other words, ontic structural realism can admit objects, as long as all there is to these objects are the relations among them. What ontic structural realism
rejects is the property-oriented metaphysics that dominates philosophy since
Aristotle: the fundamental physical objects do not have an intrinsic essence.
This is a conception of
objects that stands on its own feet, being an alternative to both the view of
objects as bare substrata and the view of objects as bundles of properties.
There is a mutual ontological dependence between objects and relations: as there
cannot be relations without objects that stand in the relations, so there cannot
be objects without relations in which they stand. Hence, if one removed the
spatial relations, there would not remain bare substrata, but there would then
be nothing (see \cite{Esfeld:2008aa, Esfeld:2011aa}). This view spells
out what \cite{Schaffer:2010ab} calls ``the internal relatedness of all things'', because certain relations -- namely the spatial ones -- are essential for the fundamental physical objects. However, this view does not end up in monism (contra \cite{Schaffer2010}): there is a
plurality of objects, namely a plurality of matter points standing in spatial relations.

Since all there is to the matter points are the spatial relations among them, the matter points are all the same. In the terminology common in physics, that is to say that they are all identical, although there is a numerical plurality of them in the universe. But their configuration is permutation invariant: a permutation of the matter points makes no ontological difference, because it does not change the spatial relations. Nonetheless, given that at least in the actual universe, the spatial configuration of matter is not symmetric, each matter point distinguishes itself from all the other ones through some of the spatial relations in which it stands. In brief, its position in the network of spatial relations distinguishes each matter point, but it is irrelevant which matter point occupies the position in question.

To conclude this section, let us come back to the term ``primitive ontology'' that is used in quantum physics to refer to the distribution of matter in physical space whose state the quantum formalism describes. We propose to endow this term with a threefold substantial meaning:

\begin{enumerate}
    \item The ontology is primitive in the sense of fundamental: the matter points are not composed of anything, but they compose everything else.

    \item It is primitive in the sense of referring to primitive objects: the
        matter points have no intrinsic properties. However, they are not
        bare substrata either. The spatial relations in which
        they stand are their essence.

    \item It is primitive in the sense of factual: the configuration of matter
        points is simply there.
\end{enumerate}

The argument for this ontology is that it is the most parsimonious way to conceive a fundamental ontology of the natural world that is able to account for the empirical data: matter
points connected by spatial relations. Consequently, all
change is change in the spatial relations. If one endorses this fundamental ontology, it is evident that calling for a primitive ontology of space and time in addition to a primitive ontology of matter is pointless: matter and space-time come in a package in a proposal for a fundamental ontology, as do geometry and dynamics in a proposal for the dynamical structure that captures the change in the configuration of matter.  

\section{Dynamical holism: Introducing dynamical structure}
\label{sec:dynamical}

Let us now consider change in the same general and parsimonious manner in which
we conceived the fundamental ontology of matter points. To be general and at the
same time precise, we cast this task in the language of mathematics.  Let
$\Omega$ denote the set of all configurations of spatial relations between $N$
matter points that can be realized in nature. In the most general case, an
element $\Delta\in\Omega$ is represented by an unordered set
$\Delta=\{\Delta_{ij}\}_{i,j=1,\ldots,N,j\neq i}$ of $N(N-1)$ spatial relations
$\Delta_{ij}$; note however that in physically relevant cases, for
instance, due to symmetry and vector space relations, $\Delta$ is usually
heavily overdetermined by specifying all relations $\Delta_{ij}$. Each
$\Delta_{ij}$ denotes the spatial relation of a matter point $j$ with a matter
point $i$, for instance in terms of distance and direction using an arbitrary
labeling. While useful in the notation, a labeling of the matter points is
artificial: as mentioned in the preceding section, the matter points are
permutation invariant.

Let us suppose that there is smooth change in the spatial relations. Hence, there is at least one possible ordering of the elements in
$\Omega$. Let us represent it by means of a smooth map
$\Delta_{(\cdot)}:\mathbb T\to \Omega$, $t\mapsto \Delta_t$ with $t$ in an
interval $\mathbb T\subseteq\mathbb
R$ being the
parametrization of this ordering. We shall refer to $t\mapsto\Delta_t$ as
\emph{motion} and to the parameter $t$ as \emph{time}. We have access to the law
of this change through the first derivative with respect to $t$, that is,
\begin{align}
    \label{eq:vel-field}
    v_t(\Delta_t) := \frac{d}{dt} \Delta_t,
    \qquad
    \text{for all }
    t\in\mathbb T.
\end{align}
This law is given in terms of a \emph{velocity field} $v_{(\cdot)}:\mathbb T\times\Omega\to T\Omega$, $(t,\Delta)\mapsto v_t(\Delta)$, where $T\Omega$ is
the tangent space
of $\Omega$, that is, the space of all possible changes of the configurations of
spatial relations in $\Omega$.

As the fundamental ontology of matter points that we propose does not imply
absolutism about space, so it does not imply absolutism about time: time derives
from change. Again, we follow Leibniz for whom time is the order of succession
-- that is, the order in the change in the universal configuration of matter
points, with that order having a direction (see notably Leibniz' third letter, §
4, and fourth letter, § 41 in \cite{Leibniz:1890aa}, pp. 363, 376, English
translation \cite{Leibniz:2000aa}). Hence, there is no time without change; but
with time, there is a unique order of the change in the universal configuration
of matter points -- in other words, a unique ordering of the elements in
$\Omega$. To put it the other way round, what makes it that this order is
temporal is that it is unique and that it has a direction.

On the one hand, if one formulates relationalism in this manner,
time is treated as emerging from the order of the configurations of matter,
provided that this order is unique and that it has a direction. On the other
hand, this relationalism clearly vindicates time: it is distinct from the
radical stance that rejects such a unique order and thereby arguably amounts to
rejecting any notion of time, since it can then not even be said that an event
$e_{2}$ lies between two other events $e_{1}$ and $e_{3}$ without specifying yet
another event $e_{4}$ as reference point. Such a radical stance is sometimes taken to be implied by the general theory of relativity (especially in its Hamiltonian formulation, as argued by \cite{Earman:2002}; but see \cite{Maudlin:2002aa} against \cite{Earman:2002}). It is notably defended by \cite{Rovelli:2004aa} (in particular sections 1.3.1, 2.4, 3.2.4) for quantum gravity; but see \cite{Gryb:2015aa} (in particular pp. 23-24, 27-29, 35-38) for an argument in favour of the view of time set out here covering also the domain of quantum gravity.

Nonetheless, although the topology of time thus is absolute, there is no unique
measure of time. That is to say, the metric of time depends on the choice of a
reference subsystem within the universe relative to which time is measured. In
fact, any subsystem can be employed to measure duration.
However, care has to be taken in the choice. Usually, the simpler the motion of
the reference subsystem with respect to the investigated subsystem is, the simpler
will be the form of the inferred velocity field $v_t$ in
\eqref{eq:vel-field}. An obvious example of a simple
reference subsystem is the circular motion of a pointer on a dial of a watch,
the arc length drawn by its pointer being directly related to $t$.

The task of physics then is to find a general law of motion that captures the
change in the spatial relations among the matter points in time $t$. A physical theory can
often be inferred on the basis of observed velocity fields \eqref{eq:vel-field}
that exemplify the general law of motion well in certain regimes. Having
obtained a possible $v_t$ by means of such a physical theory, it should also
apply to general motions $t\mapsto\Delta_t$. Hence, one can turn the definition
in \eqref{eq:vel-field} around and study all motions $t\mapsto \Delta_t$ that
fulfill
\begin{align}
    \label{eq:physics}
    \frac{d}{dt} \Delta_t = v_t(\Delta_t),
     \qquad
    \text{for all }
    t\in\mathbb R.
\end{align}
If $v_t$ is regular enough, the motion $t\mapsto\Delta_t$ -- and with it the
predictions of this physical theory  for all $t\in T$ -- are uniquely determined
by deduction from the initial condition $\Delta_0\in\Omega$. 

As mentioned in section~\ref{sec:spatial}, on the relationalism about space that we endorse, the
change in the spatial relations also determines the geometry to the following
extent: obviously, a first choice has to be made how to mathematically
represent the spatial relations $\Delta\in\Omega$, e.g., by elements in an Euclidean vector space;
otherwise, our access to the velocity field $v_t$ in \eqref{eq:vel-field} would remain
purely academic. Depending on this choice, the resulting mathematical
representation of the dynamical law in \eqref{eq:physics} found by physics may
appear sufficiently simple, in the sense that it allows for efficient
computations. If that is not the case, one may try to improve
the situation by adapting the geometric representation of the spatial relations
by exploiting, e.g., certain symmetries of the dynamical law. Such a
simplification may require the replacement of the initial geometry by, e.g.,
curved linear spaces; it may, as in the case of general relativity theory,
even involve a dynamical adaption of the geometry subject to the change in
the spatial relations among the matter points.

For our examples below, we will restrict ourselves to the special case
that permits describing the configuration of spatial relations $\Delta\in\Omega$
in terms of Euclidean vectors in $\mathbb R^3$. To arrive at such a
representation one introduces an arbitrary labeling of the $N$ matter points and
choses Euclidean arbitrary coordinate axes around a point $N$. In this coordinate
system we may then represent the spatial relations between the $N$ matter points
by a Euclidean vector $Q=(q_i)_{i=1,\ldots,N-1} \in \mathbb R^{3(N-1)}$. Here,
the vector $q_i$ relates to the spatial relation $\Delta_{iN}$. Recall,
however, that the matter points are permutation invariant and that the choice of
the reference point as well as of the axes is arbitrary. Hence, the configuration
of spatial relations $\Delta\in\Omega$ must be identified by the equivalence
class of all such $Q$ choosing different labeling and coordinate axes. Note
that, when restricting to Euclidean coordinate axes, the latter corresponds to
modding out rotations around the matter point $N$. Similarly, the
parametrization of the motion $t\mapsto\Delta_t$ can then be represented by a
map $Q_{(\cdot)}:\mathbb T \to \mathbb R^{3(N-1)}, t\mapsto Q_t$, and the
velocity by
\begin{align}
    \label{eq:equation-of-motion}
    \frac{d}{dt} Q_t = v_t(Q_t),
    \qquad
    Q_t=(q_{1,t},\ldots,q_{N-1,t})
    \quad
    \in
    \mathbb R^{3(N-1)}.
\end{align}

The change in the spatial relations among the matter points implies that a velocity field has to be specified for each transition from one universal configuration to another one. The pertinent
question then is how this can be done in such a way that specifying initial
conditions at an arbitrary time and plugging them into a law that fits into
equation (3) is sufficient to fix the motion of the matter points at \emph{any}
time. It is here that constant parameters -- such as mass, charge, total energy,
constants of nature, etc. -- or parameters at an initial time -- such as
momenta, a wave function, fields, etc. -- come into play. They are the means to
achieve this result: they are such that specifying an initial value of them and
inserting that value into a law of the type of equation (3) enables us to
capture the change in a given configuration of matter points in a manner that is
as simple and as informative as possible, obtaining as output the change in the
configuration of matter points -- that is, the velocity field -- for \emph{any}
time. Let us call the parameters that are necessary to achieve this result the
\emph{dynamical structure}. Let us consider two examples.

\paragraph{Example 1) Classical mechanics of gravitation:} For Euclidean spatial
relations, using the same notation as above to denote the motion by $\mathbb
R\ni t\mapsto
Q_t=(q_{1,t},\ldots,q_{N-1,t})\in\mathbb R^{3(N-1)}$, one may set up a
relationalist version of classical gravitation as follows. The velocity field in \eqref{eq:equation-of-motion} given as
$v_t=(v_{1,t},\ldots,v_{N-1,t})\in\mathbb R^{3(N-1)}$ is ruled by the equations
\begin{align}
    \label{eq:change-vt}
    \frac{d}{dt} v_{k,t}(Q)
    =
    -\frac{1}{M_k} \nabla_k V(Q),
    \qquad
    \text{for all } k=1,\ldots,N-1, \text{ and } Q\in\mathbb R^{3(N-1)},
\end{align}
where the map $V:\mathbb R^{3(N-1)}\to\mathbb R$ is given explicitly by
\begin{align}
    \label{eq:potential}
    V(Q) := -\frac{1}{2}\sum_{k=1}^{N-1}\sum_{j\neq k} G \frac{ m_k m_j}{|q_k-q_j|},
\end{align}
and $G, M_1,m_1,\ldots,M_{N-1},m_{N-1}\in\mathbb R^+$ are additional parameters.
Recall that $q_k$ is the spatial relation between matter point $k$ and $N$ and
that, for the geometry considered here, $q_k-q_j$ relates to $\Delta_{kj}$
as discussed above, which is the spatial relation between the respective
matter points $k$ and $j$. Together with further parameters denoted by $\dot
q_{k,0}\in\mathbb R^3$, the equations in \eqref{eq:change-vt} uniquely determine
a velocity field that is given by
\begin{align}
    \label{eq:vel-classical-mechanics}
    v_t(Q_t)
    =
    \left( v_{1,t}(Q_t),\ldots,v_{N-1,t}(Q_t) \right),
    \quad
    \text{for}
    \quad
    v_{k,t}(Q_t) := \dot q_{k,0} + \int_0^t \left(-\frac{1}{M_k} 
    \nabla_k V(Q_s)\right)\, ds.
\end{align}
These additional parameters \textendash{} that is, the relationships described
in \eqref{eq:change-vt} and \eqref{eq:potential}, Newton's constant of
gravitation $G$, inertial and gravitational masses $M_k$ and $m_k$, and the
initial velocities $\dot q_{k,0}$  \textendash{} make up the dynamical structure
of this version of classical gravitation. They are the only degrees of freedom
left in the choice of an admissible velocity field $v_t$. Thanks to
\eqref{eq:equation-of-motion}, specifying these parameters together with the
initial configuration $Q_0\in\mathbb R^{3(N-1)}$ determines the motion $t\mapsto
Q_t$ uniquely.

Relationalist versions of classical mechanics have notably been proposed by
\cite{Barbour:1977aa, Barbour:1982aa} (see \cite{Barbour:1982ab} for an overview
and \cite{Barbour:2012aa} for the recent development), \cite{Belot:1999aa} and
\cite{Huggett:2006aa}. In general, on relationalism, the only change that is
admitted is change in the spatial relations among the matter points. That is why
we can cast relationalism, as done here, in terms of arbitrarily picking out one
matter point as reference point and describe all the spatial relations, as well
as their change, with respect to that point. Note that the degree of
freedom in introducing arbitrary Euclidean coordinate axes for our
representation of $\Delta$ does not affect the definition of $V$ and hence the
dynamics, which is due to the well-known Galilei invariance of the above
dynamics. The velocity law, on the other hand, might look differently for
different choices of labeling, which is to be expected, since the dynamics can
discriminate between the $N$ matter points as we will explain in section~\ref{sec:spatial and dynamical}.
 
Consequently, on relationalism, there is no possible world with only one or two
particles that move in absolute space, and there is no such thing as inertial
motion or rotation of the configuration of matter as a whole, since this could
be only motion or rotation of the whole configuration with respect to an
absolute background space. Nonetheless, relationalism has the means to conceive
the notion of inertial motion as a particularly simple and regular motion of
some matter points relative to other matter points within a given configuration
and thereby define inertial reference frames. The same dynamics 
as above but described with respect to another reference point will for example
look more difficult as an additional force term would then appear (much like the
fictitious forces in Newtonian mechanics).

\cite{Huggett:2006aa} has put forward a proposal how this can be done on the
basis of an ontology that endorses nothing over and above Leibnizian relations
among point particles. Consider the idea of the entire history of the change in
the spatial relations among the matter points in the universe. The way in which
that change occurs manifests certain salient patterns. On this basis, one can
single out the notion of inertial motion as the idea of a particularly regular
and simple motion. One can then introduce the idea of inertial frames, formulate
the Newtonian laws of motion as the laws that hold in these frames, and conceive
acceleration as in Newtonian mechanics as deviation from that simple motion.
Thus, the absolute acceleration that one attributes to a particle in Newtonian
mechanics at a time is not an intrinsic property of that particle but a quantity
that derives from the evolution that the spatial relations in the universal
configuration of matter take. In brief, this strategy is the same as the one of
conceiving the geometry as fixed by the change in the spatial relations: in both
cases, one grants ontological priority to the dynamical evolution that a given
universal initial configuration of matter points takes and introduces the
geometry of space or space-time as well as dynamical notions such as inertial
motion and acceleration qua deviation from inertial motion on the basis of
salient patterns that manifest itself in that evolution. Huggett's proposal
makes evident that one cannot simply read an ontology off from the formalism of
a physical theory: although absolute space and time as well as absolute
acceleration figure in the physical theory of Newtonian mechanics, endorsing
this physical theory does not commit us to an ontology of absolute space and
time, since one can understand the role of these quantities in Newtonian
mechanics also on the basis of an ontology that admits only Leibnizian
relations.

However, due to the lack of an absolute reference frame, our relationalist
version of classical gravitation set out above in a precise mathematical manner
in terms of the spatial relations and their change relative to one matter point
arbitrarily picked out as reference point cannot be directly equivalent to Newton's version formulated in absolute space. Nonetheless, for any number of $N$ matter points, both theories approximate each
other in the special case of having the $N$-th matter point at a large distance
with respect to the other $N-1$ ones. This is due to the fact that in Newton's
absolute space, the acceleration of the $N$-th matter point, being
proportional to the weighted sum of inverse squared distances, is arbitrarily
close to zero provided the distance to the nearest of the other $N-1$
matter points is sufficiently large. In consequence, shifting the coordinate
system from the absolute origin to the absolute position of the $N$-th matter
point introduces only arbitrarily small fictitious forces for all other matter
points, and therefore, yields a good approximation of the
relationalist version of classical gravitation in terms of $Q_t$ defined above.

Coming back to the discussion of inertial motion, in the relationist version of classical gravitation defined above, the $N$-th
matter point can in this special case therefore serve as an almost inertial reference
frame existing in the universe.
Although this matter point is identical to all the others, the dynamics
discriminates it from the others since it conducts a particularly simple
motion with respect to them. Taking it as reference thus
results in a particularly simple representation of the velocity field $v_t$ 
as given in \eqref{eq:change-vt}. Consequently, the existence of a sufficiently distant fixed
star would make it experimentally arbitrarily hard to discriminate between
Newton's and our relationalist version of classical gravitation. Furthermore,
also in the relationist version, it may be possible to conceive reference
frames that are at least for all practical purposes inertial. One may therefore
introduce $\frac{d}{dt} v_{k,t}(Q_t)$ as the acceleration of the $k$-th matter
point and recast the dynamical structure introduced above in the spirit of
Newton's axioms of classical mechanics.

\paragraph{Example 2) Bohmian quantum mechanics:} As mentioned at the end of
section~\ref{sec:introduction}, Bohmian mechanics is the most elaborate quantum theory without observers that
admits a primitive ontology of a configuration of matter
in physical space (namely point particles as in classical mechanics) and
that formulates a dynamical law for that configuration (and not just for the
quantum state, as does the Schrödinger equation). This theory can be cast as a quantum theory of classical
gravitation (since, however, the relevance of gravitation in
the realm of light elementary particles might seem questionable -- and we use it
only for pointing out the parallels with its classical version --, one should
rather think of this theory as describing a gas of electric charges by replacing
the gravitational constant $G$ with Coulomb's constant $(4\pi\epsilon_0)^{-1}$
and the gravitational masses $m_k$ with the electric charges $e_k$). In Bohmian mechanics, the
$t$-dependence of $v_t$ is given as functional of a map $\Psi_{(\cdot)}:\mathbb
R\times \mathbb R^{3(N-1)}\to\mathbb C, (t,Q)\mapsto \Psi_t(Q)$ by 
\begin{align}
    \label{eq:BM-vel}
    v_{k,t}(Q_t) := 
    \frac{\hbar}{M_k}\Im \frac{\nabla_k\Psi_t(Q)}{\Psi_t(Q)},
\end{align}
where $\Psi_t$ is requested to be 
a square-integrable solution to the Schrödinger equation
\begin{align}
    \label{eq:schroedinger}
    i\hbar \partial_t \Psi_t(Q) 
    =
    H\Psi_t(Q),
    \qquad
    \text{for } Q\in\mathbb R^{3(N-1)},
\end{align}
for the operator
\begin{align}
    \label{eq:Hamiltonian}
    H
    =
    \sum_{k=1}^N -\frac{\hbar^2}{2M_k} \Delta_k + V(Q).
\end{align}
Here, $\hbar\in\mathbb R^+$ denotes an additional constant, while $G$, $M_k$, $m_k$, and
$V$ are the same mathematical entities as above. For a sufficiently
regular $\Psi_0$, equation \eqref{eq:schroedinger} admits
a unique solution
\begin{align}
    \Psi_t := e^{-it H}\Psi_0
\end{align}
fulfilling $\Psi_t|_{t=0}=\Psi_0$, and therefore, together with
\eqref{eq:BM-vel}, yields a unique velocity field
\begin{align}
    \label{eq:vel-bohmian-mechanics}
    v_t(Q_t)
    =
    \left( v_{1,t}(Q_t),\ldots,v_{N-1,t}(Q_t) \right)
    \qquad
    \text{for}
    \qquad
    v_{k,t}(Q_t) := 
    \frac{\hbar}{M_k}\Im \frac{\nabla_k(e^{-it H}\Psi_0)(Q)}{(e^{-it H}\Psi_0)(Q)}.
\end{align}
These additional parameters \textendash{} that is, the relationships described
in \eqref{eq:BM-vel}, \eqref{eq:schroedinger}, and \eqref{eq:potential},
Planck's constant $\hbar$, Newton's gravitational constant $G$, inertial and
gravitational masses
$M_k$ and $m_k$, and the initial wave function$\Psi_0$ \textendash{} make up the dynamical
structure. They are the only degrees of freedom left in the choice of a velocity
field $v_t$ admissible in this Bohmian setup.  Again, thanks to
\eqref{eq:equation-of-motion}, specifying these parameters together with the
initial configuration $Q_0\in\mathbb R^{3(N-1)}$ determines the motion $t\mapsto
Q_t$ uniquely.  

Note the similarity between classical mechanics and Bohmian quantum mechanics as set out here. Both theories fit into the framework of the first order
differential equation \eqref{eq:equation-of-motion}, which requires the
specification of the velocity field $v_t$. The fact that 
in classical mechanics
$v_t$ itself is defined by specifying its derivative as in \eqref{eq:change-vt},
which effectively results in a second order equation, concerns only the way of
formulating the dynamical structure but does not affect the ontology, as we shall explain in section~\ref{sec:spatial and dynamical}. All admissible candidates for $v_t$ can be found with the help of only a few additional parameters and the relations
\textendash{} cf. \eqref{eq:change-vt} and
\eqref{eq:BM-vel}-\eqref{eq:schroedinger}, respectively \textendash{}, which
rule the $t$-dependence of $v_t$ \textendash{} cf.
\eqref{eq:vel-classical-mechanics} and \eqref{eq:vel-bohmian-mechanics},
respectively. Specifying further parameters \textendash{} that is, the initial
velocities $\dot q_{k,0}$ and the initial wave function $\Psi_0$, respectively
\textendash{} uniquely determines the velocity field $v_t$, which in turn, for a
given initial configuration $Q_0$, uniquely determines the motion $t\mapsto Q_t$
as solution of \eqref{eq:equation-of-motion}.

Bohmian mechanics replaces the specification of
an initial velocity in classical mechanics with the specification of an initial
wave function. The wave function then serves to determine the velocity of the
particles, as described in equation \eqref{eq:vel-bohmian-mechanics}, whereby
the interactions between the matter points enter into the theory
via the Hamilton operator \eqref{eq:Hamiltonian}. The wave function is defined
on configuration space, that is, the mathematical space of all possible particle
configurations: each point of that space represents a possible three-dimensional
particle configuration; for $N$ particles, the dimension of the configuration
space accordingly is $3N$. On relationalism, this means that the wave function
and its evolution have to be conceived as being defined on the space of all the
possible spatial relations among the $N$ particles, which can be done as spelled
out above by arbitrarily choosing one particle as reference point as well
as coordinate axes and describing the spatial relations relative to that
particle; we note that thanks to the special form of
\eqref{eq:vel-bohmian-mechanics} it is well-known that the resulting dynamics is
invariant under rotations, and hence, under the choices of Euclidean coordinate
axes. The configuration space then is $\mathbb R^{3(N-1)}$. Again, as in
classical mechanics, Bohmian mechanics takes for granted that the evolution of
the universal configuration of matter points is such that the spatial relations
among the matter points can be represented as being embedded in Euclidean space,
that is, being expressed in terms of Euclidean lengths and angles. Inertial
reference frames can then be introduced as spelled out above for classical
mechanics. In this framework, the wave function and its evolution according to
the Schrödinger equation \eqref{eq:schroedinger} traces, via its role in the
Bohmian guiding or velocity equation \eqref{eq:vel-bohmian-mechanics}, the
evolution of the spatial relations among the particles, since the configuration
space is the space of all the possible spatial relations (instead of being the
mathematical space of all the possible particle positions in an absolute
physical space). Thus, if conceived in this manner, the dynamics of Bohmian
mechanics implies no more than a commitment to Leibnizian relations among the
particles, instead of a commitment to an absolute space into which they are
inserted.

However, as in the case of classical mechanics, so also in the case of Bohmian
quantum mechanics, our relationalist construction is not equivalent to the
formulation of the theory in terms of an absolute space. Nonetheless, as
discussed above, the one version approximates the other, in particular in the
special case of the availability of a distant fixed star that can serve as
reference point. This is also how quantum mechanics is employed in real world
applications, as coordinates are always measured with respect to a reference
point in the laboratory frame, which serves as an inertial frame with respect to
the fixed stars.

Bohmian mechanics is not the only quantum theory without observers that relies
on a primitive ontology of a configuration of matter in ordinary space and that
formulates a dynamics for that configuration by means of the quantum mechanical
wave function. One can also employ the quantum dynamics proposed by
\cite{Ghirardi:1986aa} (GRW) to do so. In GRW, the evolution of the wave
function $\Psi_t$ is given by a modified Schrödinger equation. The latter can be
defined as follows: For a rate $N\lambda$ the wave function undergoes
spontaneous jumps, while for the time intervals between two successive jumps,
$\Psi_t$ evolves according to the usual Schrödinger equation. At the time of
each jump the wave function $\Psi_t$ undergoes an instantaneous collapse
according to
\begin{align}
    \Psi_t \mapsto \frac{(L^x_k)^{1/2} \Psi_t}{\|(L^x_k)^{1/2} \Psi_t\|},
\end{align}
where the localization operator $L^x_k$ is given as a multiplication operator of
the form
\begin{align}
    L^x_k := \frac{1}{(2\pi\sigma^2)^{3/2}} e^{-\frac{1}{2\sigma^2}(x_k-x)},
\end{align}
and $x$, the center of the collapse, is a random position distributed according
to the probability density $p(x)=\|(L^x_k)^{1/2} \Psi_t\|^2$. This
modified Schrödinger evolution captures in a mathematically precise way what the
collapse postulate in textbook quantum mechanics introduces by a \emph{fiat},
namely the collapse of the wave function so that it can represent localized
objects in physical space, including in particular measurement outcomes. GRW
thereby introduce two additional parameters, the mean rate $\lambda$ as well as
the with $\sigma$ of the localization operator, which can be regarded as new
natural constants whose values can be inferred from (or at least bounded by)
experiments (e.g., chemical reactions on a photo plate, double slit experiments,
etc.). An accepted value of the mean rate $\lambda$ is of the order of
$10^{15}s^{-1}$. This value implies that the ``spontaneous localization'' process
for a single particle occurs only at astronomical times scales of the order of
$10^{15}s$, while for a macroscopic system of $N\sim10^{23}$ particles, the
collapse happens so fast that possible superpositions are resolved long before
they would be experimentally observable. Moreover, the value of $\sigma$ can be
regarded as localization width and an accepted value is of the order of
$10^-7m$. The latter is constraint by the overall energy increase of the
universe wave function that is induced into by the localization processes.

However, it is obvious that modifying the Schrödinger equation is, by itself,
not sufficient to solve the measurement problem: to do so, one has to answer the
question of what the wave function and its evolution represents. One therefore
has to add to the GRW equation a link between the evolution of the
mathematical object $\Psi_t$ in configuration space and the distribution of
matter in physical space in order to account for the outcomes of experiments
and, in general, the observable phenomena. \cite{Ghirardi:1995aa}
accomplish this task by taking the evolution of the wave function in
configuration space to represent the evolution of a matter density field in
physical space. This then constitutes what is known as the GRWm theory
and amounts to introduce in addition to $\Psi_t$ and its time evolution a field
$m_t(x)$ on physical space $\mathbb R^3$ as follows:
\begin{align}
    m_t(x) = \sum_{k=1}^N M_k \int d^3x_1\dots d^3x_N\, \delta^3(x-x_k)
    |\Psi_t(x_1,\dots,x_N)|^2.
\end{align}
This field $m_t(x)$ is to be understood as the density of matter in physical
space $\mathbb R^3$ at time $t$ (see \cite{Allori:2008aa}, section 3.1).

Hence, on this theory, there are no particles -- or, more generally speaking,
there is no plurality of fundamental physical systems. There is just one object
in the universe, namely a matter density field that stretches out throughout
space and that has varying degrees of density at different points of space, with
these degrees of density changing in time. Again, the matter density is
primitive (= propertyless) stuff, as pointed out by \cite{Allori:2013aa}:
\begin{quote} Moreover, the matter that we postulate in GRWm and whose density
    is given by the $m$ function does not ipso facto have any such properties as
    mass or charge; it can only assume various levels of density.
    (\cite{Allori:2013aa}, pp. 331--332) \end{quote} However, it remains unclear
what could constitute the difference in degrees of stuff at points of space, if
matter just is primitive stuff. The GRWm theory thus is committed to the view of
matter being a bare substratum with its being a primitive fact that this
substratum has various degrees of density at points or regions of space. In a
nutshell, there is a primitive stuff-essence of matter that furthermore admits
of different degrees of density. In comparison to this ontology, the particle
ontology -- as implemented in classical mechanics and Bohmian quantum mechanics
-- is the more parsimonious and the philosophically more convincing ontology: it
avoids the commitment to a bare substratum and to primitive degrees of density
by admitting just featureless particles that are completely characterized by the
spatial relations among them. It easily accounts for variations in the density
of matter in terms of variations in the spatial distances among the particles in
a given universal configuration of matter.

Moreover, the dynamics of Bohmian mechanics is physically more convincing than the GRW dynamics: due to the spontaneous localization of the wave function, the GRW dynamics is non-local also in configuration space. Since the evolution of the matter density in physical space is described by the evolution of the wave function in configuration space, this implies that the spontaneous localization of the wave function in configuration space signifies that some matter is delocated in physical space over a distance that can be arbitrarily big without travelling with any determinate speed across space (this is so even if the original GRW equation is replaced with an equation of a continuous spontaneous localization of the wave function in configuration space, as in \cite{Ghirardi:1990aa}). This delocalization of matter, which is not a travel with any determinate velocity, is quite a mysterious process that the GRWm dynamics asks us to countenance. In Bohmian mechanics, by contrast, as is evident from \eqref{eq:vel-bohmian-mechanics}, the motions of in principle all the particles are correlated, but the particles are never delocated across space: they always travel on continuous trajectories with a determinate velocity.

Apart from the matter density ontology, there is another ontology for the GRW dynamics available. This ontology goes back to \cite{Bell:2004aa} (ch. 22, originally published 1987): whenever there is a spontaneous localization of the wave function in configuration space, that development of the wave function in configuration space represents an event occurring at a point in physical space. These point-events are today known as flashes following \cite{Tumulka:2006aa} (p. 826). According to the GRW flash ontology (GRWf), the flashes are all there is in physical space. Hence, there only is a sparse distribution of single events. It is doubtful whether this sparse distribution is sufficient to account for macroscopic objects in terms of, as \cite{Bell:2004aa} (p. 205) put it, galaxies of such flashes (see the reservations that \cite{Maudlin:2011aa}, pp. 257-258, voices). Furthermore, as regards the dynamics, the GRWf theory asks us to countenance the appearance of flashes out of nothing (as well as their disappearance into nothing, they being ephemeral events).

In any case, the account that the original GRW theory envisages for measurement interactions does not work on the flash ontology -- in other words, this ontology covers only the spontaneous appearance and disappearance of flashes, but offers no account of interactions: on the original GRW proposal, a measurement apparatus is supposed to interact with a quantum object; since the apparatus consists of a great number of quantum objects, the entanglement of the wave function between the apparatus and the measured quantum object will be immediately reduced due to the spontaneous localization of the wave function of the apparatus. However, even if one supposes that a measurement apparatus can be conceived as a galaxy of flashes, there is on GRWf nothing with which the apparatus could interact: there is no particle that enters it, no mass density and in general no field that gets in touch with it either (even if one conceives the wave function as a field, it is a field in configuration space and not a field in physical space where the flashes are). There only is one flash (standing for what is usually supposed to be a quantum object) in its past light cone, but there is nothing left of that flash with which the apparatus could interact (see \cite{Esfeld:2014ac} for a detailed comparison of the proposals for a primitive ontology of quantum physics).

The dynamical structure that a physical theory introduces is in any case defined for the universe as a whole. To solve equation \eqref{eq:equation-of-motion} -- and the corresponding equations in classical as well as Bohmian quantum mechanics (equations (\eqref{eq:vel-classical-mechanics} and \eqref{eq:vel-bohmian-mechanics}) --, one has to put in initial data for the whole configuration of matter points. That is to say, the dynamical structure correlates in principle the motion of any matter point with the one of any other matter point in the universe. It is therefore appropriate to use the term \emph{dynamical holism}.

In classical mechanics, as is evident from equation \eqref{eq:vel-classical-mechanics}, one calculates the velocity for each matter point relative to each other matter point separately, and that velocity depends on the distance between the two matter points. Nonetheless, to obtain the correct velocity for any given matter point, one would strictly speaking have to take into account its relation to all the other matter points. For instance, as soon as there are dynamical parameters whose value is globally conserved -- such as the total energy --, there are non-local correlations in the motions of the matter points. Thus, the difference between classical and quantum mechanics by no means concerns a difference between a dynamical atomism and a dynamical holism. It is this one: in the dynamics for the primitive ontology, the quantum mechanical wave function is defined only for the configuration of matter as a whole; it correlates the motion of any matter point with in principle any other matter point \emph{without that correlation having to depend on the spatial distance between the matter points}.

One can therefore say that the dynamical holism is more obvious in quantum
mechanics than it is in classical mechanics: in quantum mechanics, there is a
single dynamical parameter -- the wave function -- that correlates the motion of
all the matter points; in classical mechanics, by contrast, one attributes
dynamical parameters to the matter points taken individually and figures out
their correlated motion pairwise, depending on the distance between them.
However, this difference does not justify regarding quantum mechanics as
endorsing the configuration space on which the wave function is defined, by
contrast to three-dimensional space, as the space in which the physical reality
is situated (see \cite{Albert:2015aa}, notably pp. 142-143, for such a view). In
both cases, as we have shown here, the theories can be construed as introducing
dynamical parameters in order to determine the change in the three-dimensional
spatial relations among matter points in such a way that specifying an initial
value of these dynamical parameters together with an initial configuration of
matter points is sufficient to fix the whole evolution of the spatial relations
among the matter points -- independently of whether this is defined for the
whole configuration at once, or pairwise for the matter points.

Again, the elegance of Bohmian mechanics is evident in this respect: if the fundamental ontology is one of a configuration of matter points that are characterized only by the spatial relations in which they stand, then a change in any one spatial relation implies in principle a change in all of them. That is why it is natural to conceive a dynamical parameter for the configuration as a whole (i.e. the wave function on configuration space with each point of that space representing a possible configuration of matter points) with the evolution of that dynamical parameter correlating the motion of any one matter point with in principle the motion of all the other ones. Thus, it is evident that what is at issue in quantum non-locality is only correlated motion; there is no need to bring in any mysterious delocation of matter across space (as in the GRWm ontology). However, as \cite{Einstein:1948aa} pointed out, if the motion of any one matter point would be correlated with the motion of any other one, it would be impossible to do physics (English translation of the relevant passage in \cite{Howard:1985aa}, pp. 187-188). Therefore, although the dynamical structure has to be conceived in the first place for the configuration of matter as a whole, describing how that configuration changes as a whole, that description has to include the means for also conceiving situations which are such that in order to capture the change in a given spatial relation, only the spatial relations to matter points in its immediate neighbourhood are relevant. Again, Bohmian mechanics accomplishes this task in an elegant manner as shown in\cite{Durr:2013aa} (ch. 5).   

\section{Spatial and dynamical structure}
\label{sec:spatial and dynamical}

The spatial structure is permutation invariant: a permutation of the matter points does not change the spatial relations among them. The dynamical structure, by contrast, is not permutation invariant. In the notation introduced in the preceding section, $\Delta\in\Omega$ does not
discriminate between matter points, but the velocity $v_t(\Delta_t)\in T\Omega$
may do so. In determining trajectories for the matter points that distinguish them, the dynamical structure sorts the matter points into different kinds of particles: some matter points move like charged particles, others like heavy or light particles, etc. so that they can be described as electrons, protons, neutrons, etc. A permutation of the particles' velocities therefore obviously leads to a different physical situation. However, it would be wrong-headed to conclude from this fact that some matter points are intrinsically electrons, others intrinsically protons, etc. In a nutshell, there is no intrinsic difference between the particles that sorts them into different particle species, but all the difference between them comes from differences in their motion (see \cite{Esfeld:2015aa} for a detailed argument). The reason is that the dynamical structure is defined on the configuration of the matter points as a whole, thus yielding a velocity field only for the whole configuration. Again, this is particularly evident in the case of the quantum mechanical wave function whose evolution is defined on configuration space, with each point of that space representing a possible configuration of matter as a whole.

Hence, even if a physical theory introduces dynamical parameters that are such that a value of them is attributed to the matter points taken individually, these parameters come in through their role in a dynamical structure that determines a velocity field for the configuration of matter points as a whole. Thus, in the case of Newtonian gravitation, the gravitational masses $m_k$ and the constant of gravitation $G$ are introduced in order to determine the potential $V$ that expresses a dynamical relation among the particles. In other words, the masses $m_k$ attributed to the particles are a constant that couples the motions of the particles to one another. By way of consequence, also as regards dynamics, it is to our mind misguided to endorse a property-oriented metaphysics. All the change that there is consists in change in the spatial relations among the matter points. To capture that change, dynamical relations are called for, which fix the manner in which the spatial relations among the matter points change. In brief, spatial structure as that what individuates the physical objects and dynamical structure as that what fixes their change is all that is required in a fundamental ontology of the natural world. 

It is therefore just a mathematical fact without ontological significance that in the classical setup,
equations \eqref{eq:equation-of-motion} and \eqref{eq:change-vt} are usually 
written down as second order equations. The same is possible for Bohmian mechanics: \cite{Bohm:1952aa} conceived the velocity law for the particles as a second order equation. Doing so does not change the role of the introduced quantities: given a $v_t$, the fundamental law is
\eqref{eq:equation-of-motion}. This law describes the change in the configuration of
spatial relations $Q_t$. The fundamental ontology that we propose consists in the spatial relations $Q_t$ and their change. The dynamical parameters that a physical theory introduces are mathematical variables employed to find a good
candidate for $v_t$. Thus, beside the other constants that have to be fixed, the
initial velocities in classical mechanics are just parameters that help to
easily find an admissible $v_t$ thanks to \eqref{eq:vel-classical-mechanics}. The role of the initial velocities is taken
over by the initial wave function in quantum theory. In the same vein, beside
the other constants, it determines the admissible $v_t$ by
\eqref{eq:vel-bohmian-mechanics}. Therefore, also the initial wave
function just is a mathematical variable that helps to carry out the task of finding a suitable candidate for $v_t$. On this view, hence, forces are not explanatory in the sense that they describe agent-like entities that push the matter points around. Consequently, from the perspective of a parsimonious fundamental ontology of matter points and change in their configuration, it makes no sense to argue that the second order formulation of Bohmian mechanics in terms of forces, including a specific quantum force, is explanatory superior to the first order formulation of \cite{Durr:2013aa} (see \cite{Belousek:2003aa} for that debate).

The same argument applies to fields. If one admits fields to the primitive
ontology in addition to the particles, thus conceiving them as some kind of
stuff that propagates in space, one runs into the same objection as the one
against absolute space: one is committed to a substance that stretches out to
infinity, far beyond where any particles are whose motion it could influence.
Moreover, conceiving fields as a material object propagating in space
usually runs into
the famous self-interaction problem: the field that a particle creates then
reacts back on the particle, resulting in a breakdown of the physical theory
that introduces the field (i.e. infinite energy at the position of the
particle). This breakdown hits classical electromagnetism as well as quantum
field theory, being the central reason why one has to help oneself to the
mathematical means of renormalization for all practical purposes in the latter
(see \cite{Feynman:1966aa}, pp. 699-700, for both these objections against
admitting fields as physical substances propagating in space). Therefore,
instead of enriching the primitive ontology, fields are better conceived as
being part and parcel of the dynamical structure that a physical theory
introduces in order to capture the change in the spatial relations among the
particles. In a nutshell, fields are a calculatory device instead of a physical
object. 

In classical electrodynamics such a programme can be made explicit in terms
of Wheeler and Feynman's formulation of electrodynamics, in which the
electromagnetic fields are simply replaced by the time-symmetric solution of the
Maxwell equations. Hence, the electromagnetic fields do not even figure in the
equation of motion as dynamical parameters but merely as a mathematical
expression depending directly on the trajectories of the charges. In this way it is
possible to define the theory of classical electrodynamics without even
mentioning fields. The situation is however more complicated in quantum field
theory in which the classical fields are elevated to operators by means of
second quantization, and with it, notions such as particle creation and
annihilation are introduced. If such notions are taken to be literally, any
attempt of introducing an ontology of persisting point particles seems futile.
However, \cite{Dirac:1934} himself outlined a possible ontology of persisting
particles for the electron sector of the standard model -- which can easily be
adapted to all other fermion sectors. According to his interpretation, the
vacuum state of quantum electrodynamics is to be interpreted as a sea of
persistent particles being distributed so ``uniformly'' that their mutual
electrodynamic interaction averages out, and therefore, individually, the
particles are ``hidden'' from our observation. What can be observed are only
disturbances of this uniformity, e.g., encoded in an initial state or caused by
interaction with another fermion sector, because in such situations, the
electrodynamic interaction fails to average out. In case only few particles make
up such a disturbance, the dynamics can be economically described by tracking
only those particles that deviate from the uniform motion together with the
respective holes they leave behind. This description can be formulated very
eloquently using creation and annihilating operators. What is created and
destroyed are however not the persistent particles that constitute the sea, but
only the ones that depart from the uniform motion in the vacuum state or relax
back to it (see \cite{Deckert:2010aa}, chs. 6 and 7). Finally, such an ontology might have also rich implications on a
cosmological scale. As the electrodynamic interaction averages out in the vacuum
state, the sea of particles proposed by Dirac literally appears as ``dark''.
Nevertheless, the sea is made up out of masses particles, and hence, it may
still interact gravitationally. In this case, Dirac's sea constitutes an
interesting candidate for \emph{dark matter} -- a track that has however not
been explored much.

In sum, we seek to capture the dynamical structure by conceiving
various dynamical parameters, as we seek to capture the spatial structure by
conceiving, for instance, an Euclidean space into which the matter points are
embedded. However, as that space does not exist as an absolute space in the
world, but is our means to represent the spatial relations among the matter
points, so the various dynamical parameters that physical theories introduce are
our means to represent the manner in which the matter points are dynamically
related so that the change in their spatial relations is fixed.

As reifying space leads into the impasse of having to answer the question of
what distinguishes matter from space mentioned in section~\ref{sec:spatial}, so upgrading
the dynamical parameters to intrinsic properties of fundamental objects leads into the
impasse of having to answer the question of how a physical object can reach out
to other objects and change their motion in virtue of properties that are
intrinsic to it. The search for a way out of this impasse then often results in
the view of there being forces or fields that literally propagate in space by
means of which one particle changes the motion of other particles. Such a view, however,
has been convincingly criticized as anthropomorphism by \cite{Russell:1912aa} among
others, and its complete breakdown is evident at the latest when it comes to quantum non-locality. By contrast, if there are only dynamical relations among the matter points, there is no problem to conceive interaction: the correlated change of their velocities is automatically given by the dynamical relations in which they stand. Thus, as mentioned at the end of section~\ref{sec:dynamical}, if the fundamental ontology is one of matter points standing in spatial relations, it is trivial that a change of any one spatial relation implies a change of in principle all the other ones. A dynamical structure then has to be conceived in the first place for the configuration as a whole, without it making sense to call for something that propagates the change of one spatial relation to the other ones.  

According to what is known today as Humean
metaphysics, the laws of nature are the axioms of the system
that strikes the best balance between being simple and being informative
in describing the distribution of matter throughout the whole of space
and time (see notably \cite{Lewis:1973aa}, pp. 72-75, and Loewer in this volume). Thus, the laws of nature supervene on or are fixed by the whole evolution of the configuration of matter, instead of these laws guiding or governing that evolution (even if they are deterministic).

Following \cite{Huggett:2006aa}, we've adopted the Humean strategy in setting out relationalism about space: the geometry of space is fixed by the manner in which the spatial relations among the matter points evolve, and so are inertial frames. We've thereby taken up the new, recent version of Humeanism according to which not only the laws, but also the dynamical parameters that a physical theory introduces supervene on the evolution of the distribution of matter in the universe, which is then characterized only in terms of the spatial configuration of the fundamental objects. As \cite{Russell:1927aa} puts it, \begin{quote} There are many possible ways of turning some things hitherto regarded as ``real'' into mere laws concerning the other things. Obviously there must be a limit to this process, or else all the things in the world will merely be each other’s washing. (\cite{Russell:1927aa}, p. 325)\end{quote}
The bedrock of the physical ontology then are, according to our proposal, the spatial relations among matter points. This again shows that one cannot read the ontology off from the formalism of a physical theory: there is no reason to subscribe to an ontological commitment to the dynamical parameters, as long as one is committed to a fundamental or primitive ontology that makes clear the nomological role that these parameters play in capturing the change in the relations in which the elements of the fundamental ontology stand.

Apart from \cite{Huggett:2006aa} about geometry and inertial frames, \cite{Hall:2009aa}, \S 5.2, has set out this strategy with respect to parameters such as mass and charge in classical mechanics. Furthermore, \cite{Miller:2014aa}, \cite{Esfeld:2014aa}, \cite{Callender:2014aa} and \cite{Bhogal:2015aa} have worked this strategy out as regards the wave function in quantum theories with a primitive ontology (such as Bohmian mechanics). In a nutshell, on Humeanism, the primitive ontology is the \textit{entire}
ontology: there is nothing over and above matter points standing in spatial
relations that change. Given the whole history of that change in the universe, the dynamical structure of the universe is fixed by or supervenes on that history. But there is nothing modal in the world, only the spatial relations among the matter points that change in a way that makes it possible for us to attribute dynamical parameters to them that allow us to conceive simple and general laws of nature. Note that this is a philosophical account of what laws of nature are: by granting nomological status to the dynamical parameters, one does not change the physical laws. One provides for an ontology of physics -- in particular cosmology -- in terms of spatial structure only, with dynamical structure supervening on the change in the spatial structure as a whole. 
       
However, the proponents of ontic structural realism usually do not adopt Humeanism. It is widespread among them to regard the dynamical structure as irreducibly modal (see notably \cite{Esfeld:2009aa} and \cite{French:2014aa}, chs. 9 and 10, for details). The view that we have sketched out in this paper is compatible also with that attitude: we make conjectures about the velocity law that captures the change in the spatial relations among the matter points by conceiving physical theories that attribute various dynamical parameters to the matter points and put these parameters in a nomological relationship, thus filling in the scheme of equation \eqref{eq:equation-of-motion} with concrete velocity laws (such as \eqref{eq:vel-classical-mechanics} and \eqref{eq:vel-bohmian-mechanics}). As one can spell these facts about the task of physical theories out in terms of Humeanism, so one can maintain that in conceiving these parameters on the basis of observed change, we seek to capture a dynamical structure that is there in the universe over and above the spatial structure. The resulting view then is a structuralist conception of the laws of nature in distinction to the Humean one: the laws are our manner to capture the dynamical structure that there is in nature (cf. \cite{Cei:2014aa}). That structure is irreducibly modal in that it constrains the change in the spatial relations among the matter points. In sum, in answering the first two questions outlined at the beginning of this paper -- what is matter and what are the laws -- we've laid stress on structures (spatial and dynamical ones), but we've remained neutral as regards the metaphysics of modality and laws of nature (see Loewer in this volume for a detailed discussion of that topic).

Proposing a universal, fundamental ontology of matter points as well as a universal velocity law -- and filling this scheme in with a precise physical theory such as classical or Bohmian quantum mechanics -- may at first glance seem sufficient to answer also the third question, namely how this apparatus explains the observable phenomena. Indeed, if the configuration of the
universe $Q_0$ and the correct velocity field $v_t$ were known, we could deduce
from equation \eqref{eq:equation-of-motion} a unique motion $t\mapsto Q_t$ of
the entire configuration. However, at best we know a
little about the initial configuration of a subsystem and nearly nothing about
the rest of the universe. Furthermore, even if we knew the types of parameters
that enter into the dynamical structure, both classical and Bohmian quantum mechanics suggest not only one but many possible velocity fields $v_t$, given that
for example the initial velocities or the initial wave function are free parameters. In brief, on the one hand, we seek for universal physical
theories and have such theories at our disposal; on the other hand, these
theories are useless as they stand when it comes to making predictions. 

In order to employ our theory about the universe to explain observable phenomena
-- e.g. the relative frequencies of measurement outcomes in experiments --, one must
replace our ignorance about the precise initial configuration and the
corresponding initial dynamical parameters by the next best thing: one has to formulate a
statistical hypothesis that together with the law of motion singles out a
probability measure, which in turn determines what will `typically' happen in the
sense of the law of large numbers. Boltzmann carried out this programme for
classical mechanics, and \cite{Durr:2013aa}, ch. 2, did so in the case
of Bohmian mechanics. The resulting predictions are then necessarily stochastic,
although the law of motion is deterministic. 

Depending on how chaotically the dynamical law reacts towards small changes in
the initial configuration and the initial dynamical parameters, the resulting
predictions for subsystems can be very precise -- as in the case of throwing
a stone on earth for which it is even worth to compute trajectories -- or quite
vague, as in the case of a coin flip, where the actual trajectories differ
so much that calculating them is not helpful anymore, and one has to reside to relative
frequencies of $1/2$ heads and $1/2$ tails. In sum, as in the case of a holistic dynamics mentioned at the end of the end of section~\ref{sec:dynamical}, so also in this case, quantum mechanics, at least in the Bohmian version, brings out more clearly a feature that is already there in classical mechanics: in the quantum case, small subsystems behave generically chaotic,
while in classical mechanics this is not generically the case. For this reason,
one resides much earlier to a probabilistic description in quantum mechanics
than in classical mechanics.

\paragraph{Acknowledgement.} We are grateful to Detlef Dürr, Shelly Goldstein, Tim Maudlin, Nino  Zanghì and all the participants of the summer school ``The Ontology of Physics'' in Saig (Black Forest, Germany) in July 2015 for discussions on the proposal set out in this paper. Andrea Oldofredi's work was supported by the Swiss National Science Foundation, grant no. 105212-149650.

\bibliographystyle{apalike}
\bibliography{references_fundont}

\end{document}